\documentclass[12pt,preprint]{aastex}
\usepackage{psfig}
\def\HI {H\kern0.1em{\sc i}} 

\def\radm {rad m$^{-2}$} 

\def\dg{$^{\circ}$}

\begin{document}
\title{~~\\ ~~\\ Faraday Rotation Measure Gradients from a Helical Magnetic Field in 3C\,273}
\shorttitle{3C\,273 RM Gradient}
\shortauthors{Zavala \& Taylor}
\author{R.T. Zavala\altaffilmark{1} and G.B. Taylor\altaffilmark{2,3}}
\email{bzavala@nofs.navy.mil; gtaylor@nrao.edu}
\altaffiltext{1}{United States Naval Observatory, 10391 W. Naval Observatory Rd., Flagstaff, AZ 86001-1149}
\altaffiltext{2}{National Radio Astronomy Observatory, P.O. Box O, Socorro, NM 87801}
\altaffiltext{3}{Kavli Institute of Particle Astrophysics and Cosmology,
Menlo Park, CA 94025}

\slugcomment{Accepted to ApJ Letters 2005 May 13.}

\begin{abstract}

Using high frequency (12-22 GHz) VLBA observations we confirm the
existence of a Faraday rotation measure gradient of $\sim$500 \radm\
mas$^{-1}$ transverse to the jet axis in the quasar 3C\,273.  The gradient
is seen in two epochs spaced roughly six months apart.  This stable
transverse rotation measure gradient is expected if a helical magnetic
field wraps around the jet.  The overall order to the magnetic field in
the inner projected 40 parsecs is consistent with a helical field.
However, we find an unexpected increase in fractional polarization along
the edges of the source, contrary to expectations. This high fractional
polarization rules out internal Faraday rotation, but is not readily
explained by a helical field.  After correcting for the rotation measure,
the intrinsic magnetic field direction in the jet of 3C\,273 changes from
parallel to nearly perpendicular to the projected jet motion at two
locations.  If a helical magnetic field causes the observed rotation
measure gradient then the synchrotron emitting electrons must be separate
from the helical field region.  The presence or absence of transverse
rotation measure gradients in other sources is also discussed.


\end{abstract}


\section{Introduction}

How jets are launched by massive black holes remains one of the
fundamental unsolved issues in astrophysics.  Many researchers have
suggested that magnetic fields are intimately involved in the
collimation of the jet and could determine which sources have
prominent jets and which do not \citep{mei01,koi02}.
\citet{bla93} urged observers to search for Faraday rotation
measure (RM) gradients transverse to the relativistic jet.  Such a gradient
is expected if a helical magnetic field wraps around the jet, and
among other effects, creates a Faraday screen.  For a source with a
jet axis in the plane of the sky a helical field wrapping around a jet
produces a maximum line-of-sight components along the jet boundary,
no net line-of-sight magnetic field component at the center. Blandford 
realized that this would produce a gradient in the observed RM 
across the jet. Projection effects for inclined jets produce an offset 
in the RM \citep{asa02} while preserving an RM gradient.

Faraday rotation measure gradients in 3C\,273 \citep{asa02} based on
5-8 GHz VLBA observations, present
intriguing evidence for one source in which we may be able to see the
effects of helically wound fields, and measure the sense of rotation
of an accreting black hole.  With the search for these RM gradients in
mind, we have reanalyzed 12-22 GHz VLBA polarimetry observations for
two epochs on 3C\,273 to obtain higher angular resolution
transverse to the jet axis, while maintaining good sensitivity to the
faint polarized emission.

Throughout this discussion, we assume H$_{0}$=71 km s$^{-1}$
Mpc$^{-1}$ \citep{wmap}, $\Omega_M$ = 0.27, and $\Omega_{\Lambda}$= 0.73.  
This gives a scale for 3C\,273 of 1 mas = 2.52 pc.

\section{Observations and Results}

The observations were made on 2000 January 27 (2000.07) and 2000 August 11 
(2000.61) with the ten element Very Long Baseline Array (VLBA)\footnote {The 
National Radio Astronomy Observatory is operated by Associated Universities,
Inc., under cooperative agreement with the National Science
Foundation.}. Data reduction and calibration for the 2000.07 epoch were 
discussed in \citet{zav01}. We re-examined the data for 2000.07 and made 
polarization angle maps using using five frequencies: 12.1, 
12.6, 14.9, 15.4 and 22.2 GHz. The 22.2 GHz maps were tapered to 
approximate the 12.1 GHz resolution, and all maps were convolved with a 
restoring beam matched to the 12.1 GHz beam. The observational setup for the 
2000.07 observation was repeated for the 2000.61 epoch. The same data reduction 
and calibration procedures of 2000.07 \citep{zav01} were applied to the 
previously unpublished 2000.61 data. A datacube of polarization position angle 
maps ordered in $\lambda^2$ was used to create the rotation measure maps. 
Pixels in the RM maps were blanked if the polarization position angle error 
was more than $\pm$ 10\dg\ at any frequency.

Rotation measure images for the two epochs are presented in Figure 1a (2000.07) 
and Figure 1b (2000.61). Between the two epochs the region immediately southwest 
of the Stokes I peak (i.e. in the projected direction of the jet) shows a change 
in RM. At 2000.07 the RM is $\approx$ 1800 \radm\, and 6 months later shows a south 
to north gradient with an RM from $-1000$ to $+1000$ \radm. 
 
Beyond 3 mas from the Stokes I peak (hereafter this region is referred to as 
the ``jet'') the RM distribution becomes smoother and does not change 
dramatically between the two epochs. The mean RM for the jet is 642 \radm\ 
for the 2000.07 epoch, and 652 \radm\ for 2000.61. An RM gradient transverse 
to the jet is apparent, and we discuss this 
in \S3.1.

In Figure 2 we show the RM corrected intrinsic B vectors for 3C\,273
under the assumption that the source is optically thin from 12-22
GHz. We simply took the Faraday corrected E vectors and rotated them by 90
degrees. This was done for the entire jet, and for the core west of the
Stokes I peak. This does not take relativistic aberration into account
\citep{lpb,lpg} which could change the orientation of the B vectors by
up to 20\dg. The jet B vectors are initially oriented roughly parallel
to the projected jet direction, then rotate to almost a N$-$S
orientation 6 mas southwest of the Stokes I peak. This changing B
vector orientation repeats itself as we move beyond 6 mas from the
Stokes I peak. For the 2000.61 epoch we detected polarization as far
as 15 mas down the jet, and the B vectors there have returned to a
direction parallel to the projected jet axis.
 
\section{The Faraday screen in 3C\,273}

In Fig. 1 we noted the change in the core RM at 3 mas southwest of the Stokes I peak. 
What causes the change in the observed RM? 1 mas corresponds to a projected 
distance of 2.5 parsecs at the redshift of 3C\,273. A region of that size is 
unlikely to change over a six month time interval. We thus reject a change in the 
physical conditions of a foreground Faraday screen as the cause of the 
changing rotation measure. A more likely explanation for the RM changes 
in the core over the 6 month time interval is a relativistically moving 
polarized jet component seen through a foreground Faraday screen. Changes in 
the observed rotation measure are caused by jet components sampling different 
sight lines as they move out from the core \citep{zav04}. The structure in the core 
is consistent with the small scale polarization structure reported by \citet{ja03}.  

The Faraday screen in front of the jet appears to be relatively constant in 
time as the RM does not significantly change over the 6 month 
interval presented here, or over three years \citep{zav01}. The jet Faraday 
screen is also spatially uniform, as opposed to the sub-structure 
seen in the core (Fig. 1).  
The more uniform Faraday screen begins at a projected distance of 
approximately 10 parsecs from the core (Fig. 1). This may be the distance 
from the central engine beyond which the magnetic field in the Faraday 
screen has become more uniform and smooth. 

\subsection{Evidence for a systematic RM Gradient}

We now examine the case for a transverse RM gradient as suggested by
\citet{bla93}.  We wish to take a slice across the broadest RM
distribution in the jet transverse to the jet motion. We use
\citet{kk} to define a jet direction of $-$112.7\dg\ at the location 
of the broadest RM distribution in Figure 1. In Figure 1 we
show the locations of the slices across the jet, with the orientation
of the slice perpendicular to the line connecting the Stokes I peak of
the component at a relative RA$ =-$3 mas, relative $\delta=-$6.5 mas
and the map center. Figure 3 shows the resulting RM of the slices at
the two epochs: 2000.07 by the dashed line and 2000.61 by the solid
line. A transverse gradient of approximately 500 \radm\ mas$^{-1}$ is
visible in Figure 3, with almost 4 beamwidths along the gradient.
This is much larger than the gradient of 35 \radm\ mas$^{-1}$ across
almost two beamwidths obtained with the coarser resolution of
\citet{asa02}.  This clearly demonstrates the advantage of working at
the highest frequencies the VLBA is capable of, consistent with enough
sensitivity to polarized emission.  The gradient expected by
\citet{bla93} is clearly present, and is suggestive of a helical field
wrapping around the jet. 

The RM is predominantly low and mostly negative
along the southern edge of the jet in 3C\,273 (ignoring a few small and
suspect patches of very high RM), even to a distance of 15 mas (38 pc) in
epoch 2000.61.  Along the northern edge of the source the RMs are
larger and predominantly positive.  This overall RM structure is
consistent with an ordered helical field, and suggests that this field
is the dominant source of the Faraday screen on projected linear 
scales out to at least $\sim$40 pc.


\subsection{An Alternative View -- Evidence for localized RM enhancements}

In Figure 4 we present the 12 GHz fractional polarization along the
slices shown in Figure 1. Along with a gradient in the RM we also see
a gradient in the fractional polarization. This indicates that the
polarization is more ordered as we proceed from south to north along
the transverse RM gradient.  Consider the simplest case of a jet in
the plane of the sky with a helical magnetic field wrapped around the
jet. The fractional polarization should be highest at the jet center,
where the RM is zero. The fractional polarization then falls off as we
approach either edge of the jet, as the RM depolarizes the underlying
emission \citep{gw66}. Severe projection effects are surely present in
3C\,273, yet it is still surprising to see that the highest RM in the
jet corresponds to high fractional polarization. Could the localized
RM {\it{and}} fractional polarization enhancements be the site of an
interaction of the jet with the ambient thermal gas which forms the
Faraday screen \citep{bick03}? In this case the rotation of the B
vectors in Figure 2 to approximately perpendicular to the jet motions
seen by \citet{kk} could be a compression and enhancement of the
local magnetic field due to a collision with the Faraday rotating
thermal gas. If such an interaction is responsible for the observed
RM and fractional polarization gradients, then the lack of free-free
absorption of the jet of 3C\,273 may set interesting limits on the
electron density of the thermal gas \citep{zav04}.

\subsection{RM distributions in other sources}
  
Sources with jets broad enough to be well resolved are rare
\citep{pol03}, but a few examples do exist. The jet of M87 is
resolved, but the RM data cover a relatively small area
\citep{zav02}. There is no obvious RM gradient, but the sign change of
M87's rotation measure is consistent with a helical magnetic field.
\citet{gab04} report rotation measure gradients in 4 BL Lac sources
with an angular resolution comparable to \citet{asa02}. 3C\,147, a
compact steep spectrum (CSS) quasar, has a relatively broad jet with
an RM gradient observed at 8 GHz \citep{zh04}. The quasars B1611+343
and B2251+158 are well resolved, but there is no evidence of a
transverse RM gradient \citep{zav03}.  BL Lac is marginally resolved
and no gradient is apparent \citep{zav03}.  Another quasar with a
resolved jet, B0736+017, fails to display an obvious transverse RM
gradient. At present we can say that transverse RM gradients do exist,
but these gradients are not a universal feature.

The very high fractional polarization we observe rules out internal
Faraday rotation as a cause for the observed RM. Internal Faraday
rotation is expected to cause severe depolarization, and
discontinuities in the observed position angle versus $\lambda^{2}$
plots \citep{burn,trib}. If a helical field is responsible for the
observed RM gradient, this implies a segregation of the synchrotron
emitting electrons from the helical field region.

\section{Summary}

We confirm the presence of a transverse rotation measure gradient of
500 \radm\ mas$^{-1}$, significantly larger than the gradient observed
by \citet{asa02}. This rotation measure gradient, as first suggested
by \citet{bla93}, is expected if a helical magnetic field wraps around
the relativistic jet of 3C\,273.  This field imposes an order on the
Faraday screen which we see as predominantly negative RMs along the
south edge and positive RMs along the north edge in the inner
projected 40 parsecs of the jet.  The increase in fractional
polarization along the RM gradient, and turning of the B field vectors
from parallel to perpendicular to the projected jet motion, suggest
that an interaction may also explain the gradient, although in this
case the overall order of the RM distribution has to be explained as a
coincidence, or some other mechanism that imposes an overall order to
the surrounding magnetic fields. The magnetic field responsible for
the Faraday rotation appears to have structure on sub-parsec scales
within 3 milliarcseconds (8 pc) from the core of 3C\,273. The relatively
smooth RM structure of the jet implies a well ordered (on parsec
scales) magnetic field in the Faraday screen.

The larger RM gradient and the gradient in fractional polarization
structure in 3C\,273 revealed by these observations illustrate the
advantages of observing broad jets at the highest frequencies at which
the VLBA operates.  The superb resolution and polarization sensitivity
of the VLBA from 12 -- 22 GHz enables high spatial resolution tests of
theoretical predictions which are key to understanding the physics of
relativistic jets. A convincing test of the helical magnetic field
model awaits a clear predominance of RM gradients with the expected
helicity in many sources, or the detection of a polarized counter-jet
with gradients in opposed directions \citep{asa02}.  If a helical
magnetic field is responsible for the observed RM gradients then because the
high fractional polarization observed rules out internal Faraday
rotation, the synchrotron emitting electrons must somehow be
segregated from the helical B field region.

\acknowledgments 

We thank Phil Hardee for helpful discussions which improved the paper.
RTZ thanks Alexandra Zavala for assistance with the submission.
The National Science Foundation, through its support of NRAO, has funded 
this work. This research has made use of NASA's ADS 
Abstract Service and the NASA/IPAC Extragalactic Database (NED) which is 
operated by the Jet Propulsion Laboratory, Caltech, under contract with 
NASA.

\clearpage

 
 
 
 

\begin{figure}
\epsscale{0.50}
\plotone{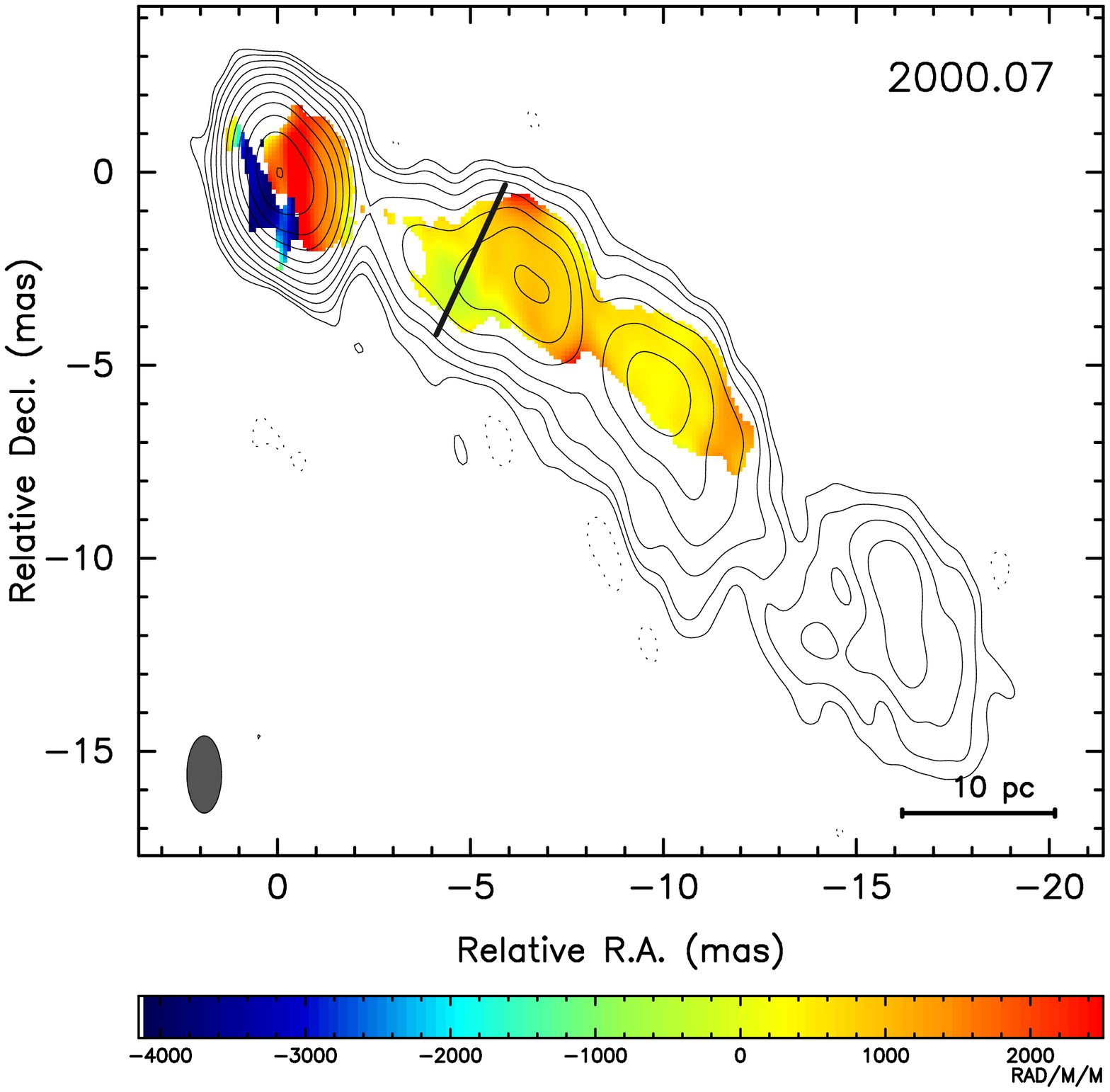}\\
~~\\
~~\\
~~\\
\plotone{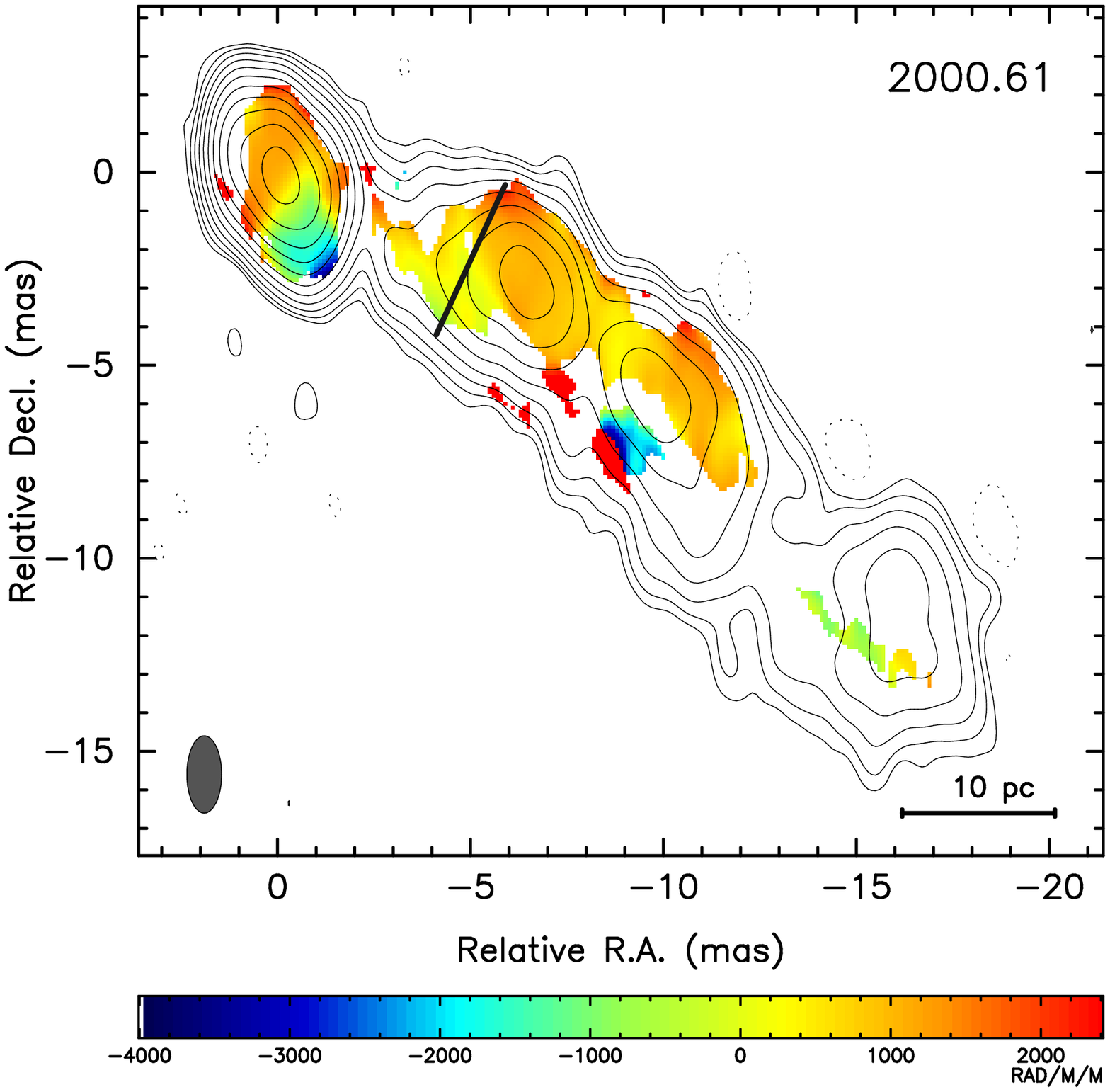}
\caption{ 
Rotation Measure image for 3C\,273 on two epochs.  Total intensity 
contours at 12.1 GHz are overlaid with contour levels starting at
12 mJy/beam and 6.6 mJy/beam for epoch 2000.07 and 2000.61 respectively and
increase by factors of 2.  The synthesized beam is drawn in the bottom-left 
corner and has dimensions of 2.0 $\times$ 0.9 mas at 0$^\circ$ for 
both epochs.  The line indicates the location of the slices shown 
in Fig.~3.
  }
\end{figure}
\clearpage

\begin{figure}
\epsscale{0.50}
\plotone{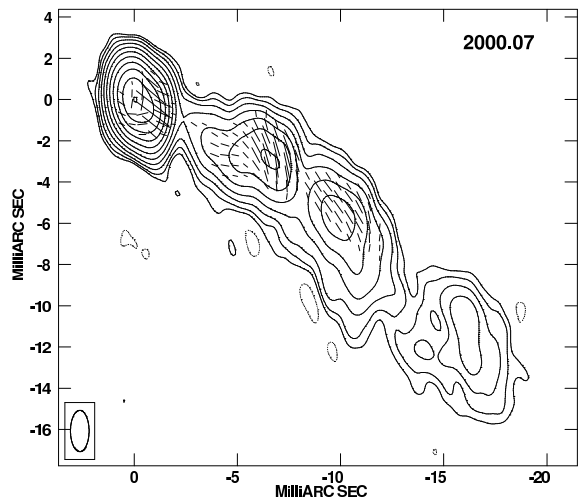}\\
~~\\
~~\\
~~\\
\plotone{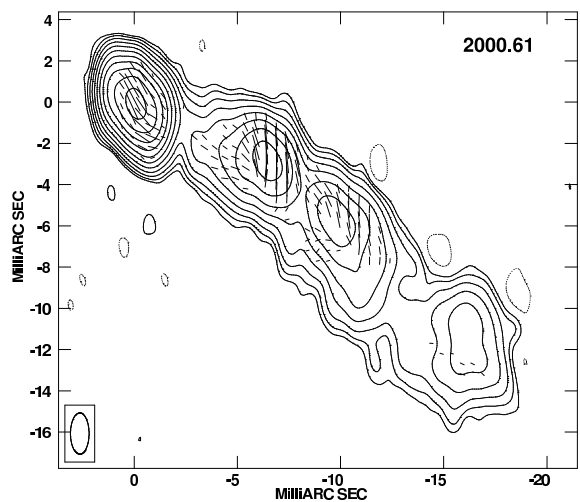}
\caption{ 
Rotation Measure corrected magnetic vectors for 3C\,273 for our two epochs 
overlaid on total intensity contours. Contour intervals are the same as for 
Figures 1a and b. Lengths of magnetic vectors are proportional to the polarized 
intensity with scales of 1 mas = 100 mJy beam$^{-1}$ for both epochs. 
}
\end{figure}
\clearpage

\begin{figure}
\epsscale{0.54}
\plotone{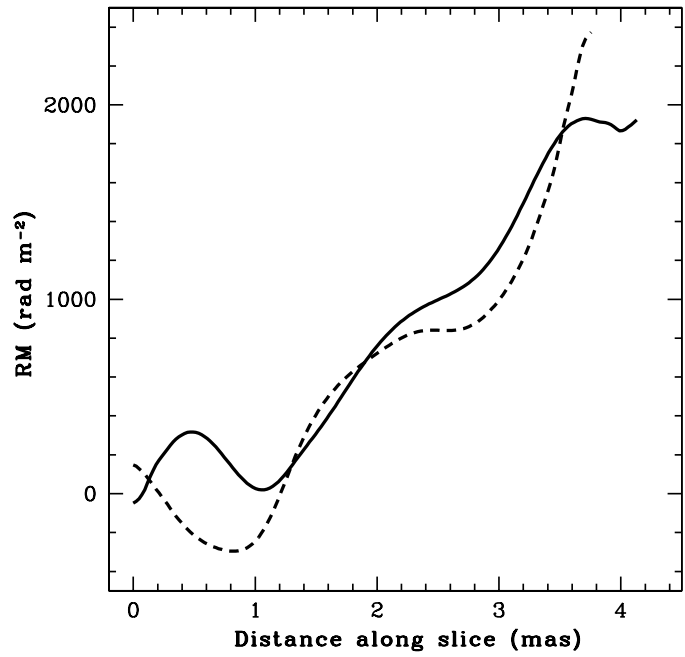}
\caption{ 
Rotation Measure image slices perpendicular to the jet axis for 
both epoch 2000.07 (dashed) and 2000.61 (solid).  The region of
the slice is indicated by the solid line in Fig.~1.
  }
\end{figure}
\clearpage

\begin{figure}
\epsscale{0.54}
\plotone{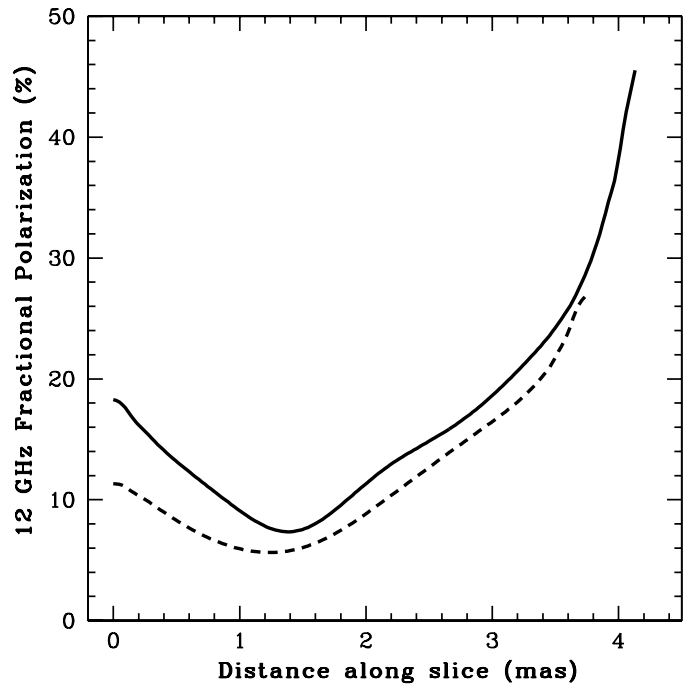}
\caption{ 
Fractional polarization slices perpendicular to the jet axis for 
both epoch 2000.07 (dashed) and 2000.61 (solid).  The region of
the slice is indicated by the solid line in Fig.~1.
  }
\end{figure}
\clearpage


\clearpage

\begin{thebibliography}{}
\bibitem[Asada et al.(2002)]{asa02} Asada, K., et al.\ 2002, PASJ, 54, L39
\bibitem[Attridge et al.(2003)]{ja03} 
Attridge, J.~M., et al.\ 2003, American Astronomical Society Meeting, 203
\bibitem[Bicknell et al.(2003)]{bick03} Bicknell, G.~V., et al.\ 2003, New 
Astronomy Review, 47, 537 
\bibitem[Blandford (1993)]{bla93} Blandford, R.~D.\ 1993, in 
Space Telescope Science Inst. Symp. Ser. 6,{\it Astrophysical Jets},
ed. D. Burgarella, M. Livio \& C.~P. O'Dea
(Cambridge: Cambridge University Press), p. 15.
\bibitem[Burn(1966)]{burn} Burn, B.~J.\ 1966, \mnras, 133, 67 
\bibitem[Gabuzda, Murray, \& Cronin(2004)]{gab04} Gabuzda, 
D.~C., Murray, {\' E}., \& Cronin, P.\ 2004, \mnras, 351, L89
\bibitem[Gardner \& Whiteoak(1966)]{gw66} Gardner, F.~F.~\& 
Whiteoak, J.~B.\ 1966, \araa, 4, 245
\bibitem[Kellermann et al.(2004)]{kk} Kellermann, K.~I., 
et al.\ 2004, \apj, 609, 539
\bibitem[Koide et al.(2002)]{koi02} Koide, et al.\ 2002, Science, 295, 1688
\bibitem[Lyutikov, Pariev, \& Blandford(2003)]{lpb} 
Lyutikov, M., Pariev, V.~I., \& Blandford, R.~D.\ 2003, \apj, 597, 998
\bibitem[Lyutikov, Pariev \& Gabuzda(2005)]{lpg} Lyutikov, M., Pariev, 
V., \& Gabuzda, D.\ 2005, Memorie della Societa Astronomica Italiana, 76, 
114 
\bibitem[Meier, Koide, \& Uchida(2001)]{mei01} Meier, D.~L., 
Koide, S., \& Uchida, Y.\ 2001, Science, 291, 84 
\bibitem[Pollack, Taylor, \& Zavala(2003)]{pol03} Pollack, 
L.~K., Taylor, G.~B., \& Zavala, R.~T.\ 2003, \apj, 589, 733
\bibitem[Spergel et al.(2003)]{wmap} Spergel, D.~N., et al.\ 
2003, \apjs, 148, 175 
\bibitem[Tribble(1991)]{trib} Tribble, P.~C.\ 1991, \mnras, 
250, 726 
\bibitem[Zavala \& Taylor (2001)]{zav01} Zavala, R.~T. \& Taylor,
  G.~B. 2001, \apj, 550, L147
\bibitem[Zavala \& Taylor (2002)]{zav02} Zavala, R.~T. \& Taylor,
  G.~B. 2002, \apj, 566, L9
\bibitem[Zavala \& Taylor (2003)]{zav03} Zavala, R.~T. \& Taylor,
  G.~B. 2003, \apj, 589, 126
\bibitem[Zavala \& Taylor (2004)]{zav04} Zavala, R.~T. \& Taylor,
  G.~B. 2004, \apj, 612, 749
\bibitem[Zhang et al.(2004)]{zh04} Zhang, H.~Y., et al.\ 2004, \aap, 415, 477 

\end{thebibliography}
\end{document}